\documentclass{svproc}

\usepackage{url}
\usepackage{amsmath}
\usepackage{graphicx}
\usepackage{array}
\usepackage{algorithm}
\usepackage{algorithmic}
\usepackage{float}
\usepackage{adjustbox}
\usepackage{booktabs}
\usepackage{subcaption}
\usepackage[utf8]{inputenc}
\usepackage{caption}
\usepackage{booktabs} 
\usepackage{verbatim}
\usepackage{comment}
\usepackage{xcolor}
\usepackage{textcomp}
\usepackage{pifont}
\usepackage{fontawesome}
\usepackage{enumitem}
\usepackage{marvosym}

\setlist[itemize]{label=\textbullet} 

\newcommand{\PreserveBackslash}[1]{\let\temp=\\#1\let\\=\temp}
\usepackage{multirow}
\usepackage{amssymb}
\newcolumntype{C}[1]{>{\PreserveBackslash\centering}p{#1}}
\newcolumntype{R}[1]{>{\PreserveBackslash\raggedleft}p{#1}}
\newcolumntype{L}[1]{>{\PreserveBackslash\raggedright}p{#1}}

\begin{document}
\mainmatter              
\title{FLTG: Byzantine-Robust Federated Learning via Angle-Based Defense and Non-IID-Aware Weighting}



%
\titlerunning{Byzantine-Robust Federated Learning }  

\author{Yanhua Wen\inst{1} \and Lu Ai\inst{1} \and Gang Liu\inst{2}\textsuperscript{\Letter} \and Chuang Li\inst{1} \and Jianhao Wei\inst{1}}
\authorrunning{Yanhua Wen et al.} 
%
\tocauthor{Yanhua Wen, Lu Ai, Gang Liu, Chuang Li, Jianhao Wei}
\institute{Hunan University of Technology and Business and also with Xiangjiang Laboratory, Hunan 410205, China\\
\email{yanhua-wen@hutb.edu.cn}, \email{1392054016@qq.com}, \email{chuangli@hutb.edu.cn}
\email{jianhao@hutb.edu.cn}
\and
Shenzhen Institute for Advanced Study, University of Electronic Science and Technology of China, Shenzhen 518055, China\\
\email{gliu29@uestc.edu.cn}}

\maketitle              

\begin{abstract}  
Byzantine attacks during model aggregation in Federated Learning (FL) threaten training integrity by manipulating malicious clients' updates. Existing methods struggle with limited robustness under high malicious client ratios and sensitivity to non-i.i.d. data, leading to degraded accuracy. To address this, we propose FLTG, a novel aggregation algorithm integrating angle-based defense and dynamic reference selection. FLTG first filters clients via ReLU-clipped cosine similarity, leveraging a server-side clean dataset to exclude misaligned updates. It then dynamically selects a reference client based on the prior global model to mitigate non-i.i.d. bias, assigns aggregation weights inversely proportional to angular deviations, and normalizes update magnitudes to suppress malicious scaling. Evaluations across datasets of varying complexity under five classic attacks demonstrate FLTG’s superiority over state-of-art methods under extreme bias  scenarios and sustains robustness with higher proportion(over 50\%) malicious clients.
\keywords{Federated Learning; Byzantine Robustness; Cosine Similarity Aggregation; Non-IID Data; Model Poisoning Attacks}
\end{abstract}
\section{Introduction}

Federated Learning(FL)\cite{konevcny2016federated},\cite{mcmahan2017communication} is a decentralized machine learning \cite{he2016deep} approach that enables clients to collaboratively train a model without exposing their private data. Each client trains a local model using its own data and sends only model updates (e.g., gradients)\cite{zeng2024bsr} to a central server for aggregation. This framework ensures data privacy by keeping sensitive  original data local, while still benefiting from the collective knowledge of all participants. FL is particularly well-suited for applications involving mobile devices\cite{gong2024towards}, blockchain \cite{2025BRFL}, and healthcare\cite{2025FedCCW}, where data security is paramount.

However, federated learning (FL) still encounters a number of challenges when applied in real-world scenarios. Its decentralized structure makes it susceptible to malicious disruptions, particularly during the model sharing phases, which can expose it to Byzantine attacks. Among the primary threats are data poisoning attacks\cite{2012Poisoning} and local model poisoning attacks\cite{2020Local},\cite{2018How},\cite{2021FLTrust}, both of which can severely affect the global model's training performance, jeopardizing the security and stability of the entire FL framework.
To tackle this security challenge, various Byzantine-robust FL approaches have been proposed\cite{2025FedCCW},\cite{2021FLTrust}, \cite{2017Machine},\cite{2018Byzantine},\cite{mai2025rflpa},\\\cite{bao2024boba},\cite{2021FLOD}, \cite{miao2024efficient}, \cite{2020Model}. The majority of these methods detect malicious clients by analyzing discrepancies in model updates, either through comparisons with other client models or the server model. These solutions face two critical limitations: (1) sensitivity to non-IID data -- conventional similarity metrics (e.g., Euclidean distance in Krum, coordinate-wise trimming in Trim-mean) fail to disentangle malicious updates from legitimate yet divergent updates caused by data heterogeneity, leading to over-filtering and model bias; (2) fragility under high malicious ratios -- most methods degrade sharply when malicious clients dominate, as static reference models or fixed thresholds become ineffective against adaptive poisoning strategies.

To address the aforementioned challenge, based on the two main categories of defense aggregation methods in federated learning, we select angle-based defense techniques to detect malicious clients. Compared with angle-based approaches, distance-based methods have the following disadvantages: First, distance-based methods typically require strict assumptions, such as knowing the number of malicious clients in advance, which are hard to meet in real federated learning environments. Second, computing Euclidean or Mahalanobis distance requires extensive mathematical operations on model parameters, which becomes computationally expensive, especially for high-dimensional models. Third, distance-based approaches are generally more effective at detecting attacks that manipulate update magnitudes, whereas angle-based methods are better suited for detecting directional anomalies in gradients. In our study, we not only focus on the direction of the gradients but also normalize the gradients to account for their magnitudes. This combination gives our method a robustness advantage that cannot be achieved by distance-based methods.

Based on the advantages of the aforementioned angle-based techniques, we propose a Byzantine-Robust aggregation algorithm, FLTG. First, we compute the cosine similarity between each client’s model update and the server’s model update. Next, we perform additional operations on the subset of clients whose cosine similarity is greater than zero. To reduce the effects of non-iid (non-independent and identically distributed) data, we identify the client within this subset that exhibits the largest angular difference from the global model in the previous round. We then calculate the cosine similarity between this reference client and the rest of the clients in the subset. We hypothesize that clients more similar to the reference client should have lower aggregation weights. Subsequently, we normalize the model updates of the client subset, scaling them to align with the server’s model update on the same hypersphere in the vector space. This step helps mitigate the influence of malicious model updates with large deviations. Finally, FLTG computes a weighted average of the normalized model updates to complete the aggregation. In conclusion, the main contributions of our work are as follows.

\begin{itemize}
    \item \textbf{Innovative Byzantine-Robust Aggregation Mechanism.}
     We integrate ReLU-based cosine similarity screening strategy and dynamic reference selection strategy driven by historical global model updates, ensuring adaptive defense against evolving attacks while minimizing over-filtering of legitimate non-IID updates.
    \item \textbf{Enhanced Robustness to Non-IID Scenario.}
     We introduce a global model-guided scoring system,  distinguishing malicious deviations from benign data heterogeneity. The inverse-proportional weighting of angular differences mitigates root dataset bias and client non-IID impacts.
    \item \textbf{Comprehensive Resilience Validation.}
    We rigorously evaluate FLTG across diverse datasets (MNIST, CIFAR-10), model architectures (CNN, ResNet20), and six attack scenarios (e.g., label-flipping, Krum attack). FLTG sustains robustness even with $>$50\% malicious clients, outperforming state-of-the-art methods in accuracy and attack resistance.
\end{itemize}

\section{Background and Related Works}
\subsection{Federated Learning}\label{sec:federated_learning}
\begin{figure}[!ht]
    \centering
    \includegraphics[width=0.8\textwidth]{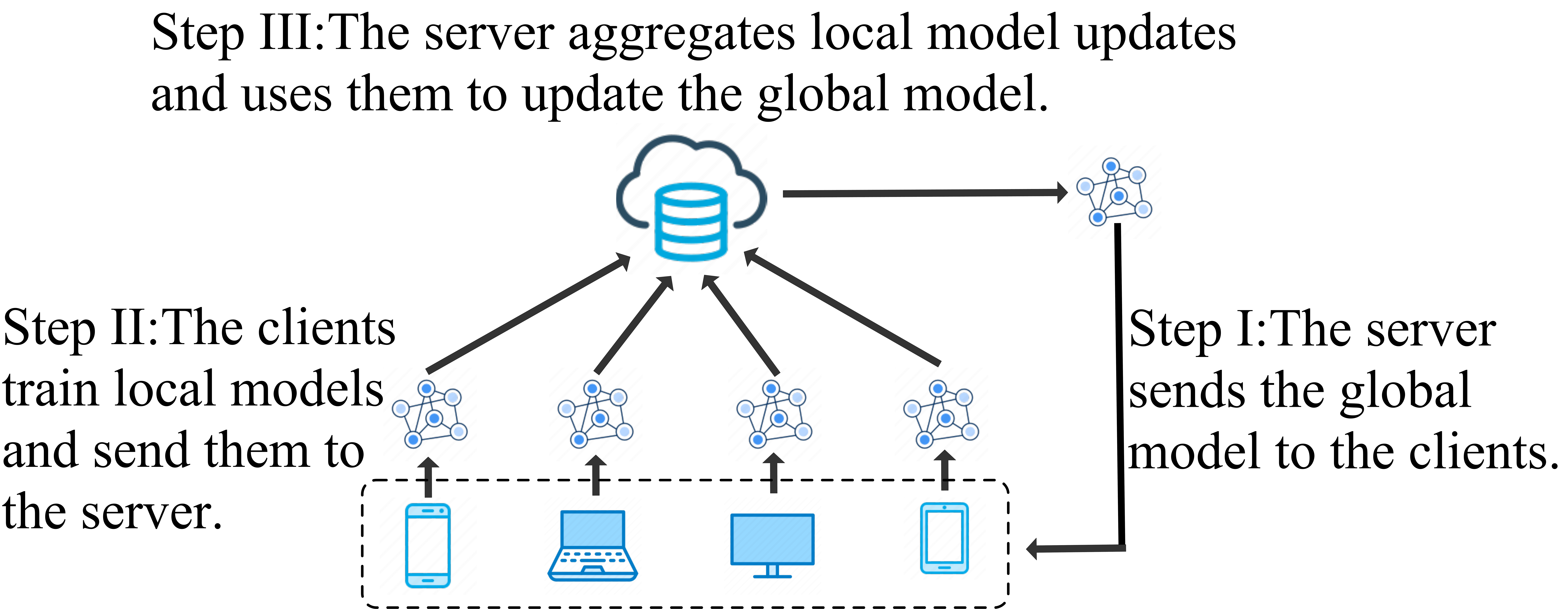}
    \caption{Federated Learning Framework.}
    \label{figures1}
\end{figure}

Federated Learning (FL) is a decentralized machine learning method where multiple clients collaboratively train a shared model without directly exchanging their data. Each client trains a local model using its own data and only shares model updates (such as gradients) with a central server. This approach ensures data privacy while enabling the construction of a global model. The implementation of Federated Learning follows these three steps (as shown in Fig.~\ref{figures1}): The server sends the global model to clients, who then train local models and send updates back. The server aggregates these updates to update the global model.

\subsection{Byzantine-Robust Federated Learning}
Byzantine-Robust Federated Learning (BR-FL) aims to maintain the reliability and integrity of model aggregation in the presence of malicious participants. While privacy-preserving techniques in Federated Learning (FL) focus mainly on safeguarding sensitive data, BR-FL addresses the challenge of ensuring the accuracy and trustworthiness of models despite adversarial attacks, such as Byzantine failures, where participants may submit corrupted or misleading updates to disrupt the training process. To combat this issue, a variety of Byzantine-robust aggregation strategies have been proposed. Aggregation rules are vital in Federated Learning, as they directly influence the system's ability to effectively mitigate malicious behavior. In the following section, we will introduce several classical aggregation algorithms.

\subsection{Byzantine-Robust Aggregation Rules}
\textbf{Krum}\cite{2017Machine}:
Krum is designed to assess the similarity between each client's model update and those of other clients using the squared Euclidean distance. The server evaluates each update by computing the sum of squared distances to its \( n - f - 2 \) closest neighbors, where \( n \) is the total number of clients and \( f \) represents the number of malicious clients. The update with the smallest sum of distances is selected as the global model update, thereby reducing the influence of malicious clients. Krum has been proven to tolerate up to \( f \) malicious clients, provided that the condition \( n > 2f + 2 \) is met. The aggregation process for Krum is formulated as follows:  
\begin{equation}
\boldsymbol{g} = \boldsymbol{g}_{i^*}, \quad i^* = \arg \min_{i \in \{1, \dots, n\}} \sum_{j \in \mathcal{N}(i)} \| \boldsymbol{g}_i - \boldsymbol{g}_j \|_2^2,
\end{equation}  

where \( \boldsymbol{g}_i \) and \( \boldsymbol{g}_j \) denote the model updates from clients \( i \) and \( j \), respectively, and \( \mathcal{N}(i) \) represents the set of \( n - f - 2 \) closest neighbors of \( \boldsymbol{g}_i \), determined based on the squared Euclidean distance.

\textbf{Trim-mean}\cite{2018Byzantine}:
Trimmed-Mean is a coordinate-wise aggregation method that enhances robustness by filtering out extreme values. Specifically, for each parameter, it discards the \( k \) largest and \( k \) smallest values before computing the mean of the remaining ones. The trim parameter \( k \), which determines the number of values removed, must be at least equal to the number of malicious clients to ensure resilience against attacks. This approach allows Trim-Mean to tolerate up to \( k < \frac{n}{2} \) malicious clients, meaning it remains effective as long as fewer than 50\% of participants are adversarial. The aggregation process follows this formulation:  
\begin{equation}
\boldsymbol{g} = \left[ \boldsymbol{g}_1, \boldsymbol{g}_2, \dots, \boldsymbol{g}_d \right],\quad \boldsymbol{g}_j = \frac{1}{n - 2k} \sum_{i \in S_j} \boldsymbol{g}_{ij},
\end{equation}  

where \( \boldsymbol{g}_{ij} \) denotes the \( j \)-th parameter in the \( i \)-th client’s model update, and \( S_j \) represents the set of remaining values for the \( j \)-th parameter after removing the \( k \) highest and lowest values. Here, \( n \) refers to the total number of clients. The final aggregated update \( \boldsymbol{g} \) is composed of all parameter values \( \boldsymbol{g}_1, \boldsymbol{g}_2, \dots, \boldsymbol{g}_d \), where \( d \) represents the total number of parameters in the model.

\textbf{Median}\cite{2018Byzantine}:
The Median aggregation method operates in a coordinate-wise manner, determining the global model update by computing the median value of each parameter across all client updates. This approach enhances robustness by mitigating the influence of extreme values. The aggregation process is mathematically defined as follows:  
\begin{equation}
\boldsymbol{g} = \left[ \boldsymbol{g}_1, \boldsymbol{g}_2, \dots, \boldsymbol{g}_d \right],\quad \boldsymbol{g}_j=\text{median}\left( \{\boldsymbol{g}_{ij} : i \in [n] \} \right),
\end{equation}  

where \( \boldsymbol{g}_{ij} \) denotes the \( j \)-th parameter in the model update from client \( i \), and \( n \) represents the total number of participating clients. The final aggregated update \( \boldsymbol{g} \) consists of all parameter values \( \boldsymbol{g}_1, \boldsymbol{g}_2, \dots, \boldsymbol{g}_d \), with \( d \) being the total number of model parameters.

\textbf{FLTrust}\cite{2021FLTrust}:
FLTrust introduces a defense mechanism in which the server manually collects a small, trustworthy dataset—referred to as the root dataset that remains untainted by poisoning attacks. Using this dataset, the server trains a reference model, known as the server model, analogous to how each client trains its local model. To assess the reliability of client updates, FLTrust computes a trust score (TS) for each client based on the cosine similarity between the client's model update and the server model update. These trust scores serve as weights during the aggregation process. Additionally, FLTrust reduces the impact of malicious updates by normalizing each client’s model update, adjusting its magnitude to align with that of the server model update. The aggregation process follows the formula:  
\begin{equation}
\boldsymbol{g} = \frac{1}{\sum_{j=1}^n TS_j} \sum_{i=1}^n TS_i \cdot \bar{\boldsymbol{g}}_i,
\end{equation}  
where \( TS_j \) represents the trust score assigned to client \( j \), and \( \bar{\boldsymbol{g}}_i \) denotes the normalized model update of client \( i \).

\section{Problem Setup}
\subsection{Attack Model}  
In federated learning (FL), adversaries can either compromise existing clients or introduce fake ones to launch attacks. These malicious clients generate corrupted local updates with the intent of manipulating the global model, a strategy known as Byzantine attacks\cite{2020Local},\cite{2021FLTrust},\cite{2018How}. The goal of such attacks is to disrupt the collaborative training process and degrade the global model’s performance. Consistent with prior work on poisoning attacks, we classify adversaries based on their level of knowledge. In the partial knowledge setting, attackers have access to their own local training data, model updates, loss functions, learning rates, batch sizes, and other related parameters. In the full knowledge setting, adversaries have complete visibility into the entire FL training process, including local training data, model updates across iterations, aggregation rules, and all relevant parameters.  

To evaluate the robustness of our method, we conduct experiments under the full knowledge setting. In this scenario, we assume the server remains uncompromised and does not collaborate with any clients. However, the server operates under an honest-but-curious model: it strictly follows the aggregation protocol without selectively discarding updates but may attempt to infer private client information by analyzing both local and global model updates during training rounds.  The main types of threats considered include:  
\begin{itemize}  
    \item Data Poisoning Attacks \cite{2021FLTrust}: Adversaries alter training labels without changing input features, such as in label flipping attacks where malicious clients' labels are changed to a target class or reassigned randomly. Detecting these attacks is difficult as the server cannot directly access client data.
    \item Model Poisoning Attacks \cite{2020Local}: These attacks modify local model updates strategically. If the aggregation mechanism doesn't filter them out, it can lead to significant performance loss or failure of the global model to converge.
    \item Backdoor Attacks \cite{2018How}: These attacks embed hidden triggers in training data. Models trained on this data will misclassify inputs containing the trigger, producing the attacker's intended output. For instance, in image classification, an attacker might insert a subtle pattern in images and mislabel them. The model will only be compromised when it encounters this pattern during inference. 
\end{itemize}

\subsection{Design Goals}  
We aim to develop a Byzantine-robust federated learning framework that focuses on server-side defense mechanisms. The framework is designed to ensure fidelity, meaning the global model's accuracy remains uncompromised in the absence of malicious clients. Simultaneously, it guarantees robustness, maintaining high global model accuracy even under adversarial attacks. To achieve these objectives, our proposed framework must be effective across different datasets, attack strategies, and scenarios with highly heterogeneous data distributions. Furthermore, it should remain functional even when the fraction of malicious clients exceeds 50\%.

\subsection{System Model}  
Our system architecture, consistent with FLTrust\cite{2021FLTrust}, involves the server maintaining a small, clean dataset (referred to as the root dataset) and following the standard three-step federated learning process outlined in Section~\ref{sec:federated_learning}. However, a key distinction exists in the second step: while clients train their local models, the server also trains a model on the root dataset, known as the server model. This additional step enables the server to utilize its own model update as a trusted reference during aggregation. Specifically, in the third step, the server compares client model updates against the server model update, leveraging directional similarity to detect and mitigate malicious contributions.  For a detailed breakdown of the workflow, please refer to Algorithm~\ref{alg:fl_framework}.

\begin{algorithm}[H] 
\caption{Federated Learning Framework}
\label{alg:fl_framework}
\begin{algorithmic}[1]
\REQUIRE A set of $n$ clients $S = \{C_1, C_2, \ldots, C_n\}$, with their corresponding training datasets $\{D_1, D_2, \ldots, D_n\}$, server's root dataset $D_0$, global learning rate $\alpha$, local learning rate $\beta$, total global rounds $R$, $\eta$ clients selected per round, local epochs $e$, and batch size $b$.
\ENSURE Global model $G$.
\STATE $G \gets$ random initialization.
\FOR{$t = 1, 2, \ldots, R$}
    \STATE // Step I: The server sends the global model to clients.
    \STATE The server randomly samples $\eta$ clients $C_1, C_2, \ldots, C_\eta$ from $\{1, 2, \ldots, n\}$ and sends $G$ to them.
    \STATE // Step II: Training local models and server model.
    \STATE // Client side.
    \FORALL{$i = C_1, C_2, \ldots, C_\eta$}
        \STATE $g_i^t = \text{ModelUpdate}(G^t, D_i, b, \beta, e)$.
        \STATE Send $g_i^t$ to the server.
    \ENDFOR
    \STATE // Server side.
    \STATE $g_0^t = \text{ModelUpdate}(G^t, D_0, b, \beta, e)$.
    \STATE // Step III: Updating the global model via aggregating the local model updates.
    \vspace{-2.8ex}
    \FORALL{$i = C_1, C_2, \ldots, C_\eta$}
        \STATE $Score_i, \bar{g}_i^t \gets \text{FLTG}$.
    \ENDFOR
    \STATE $g^t = \frac{1}{\sum_{j=1}^\eta Score_{C_j}} \sum_{j=1}^\eta Score_{C_j} \cdot \bar{g}_{C_j}^t$.
    \STATE $G^t = G^{t-1} + \alpha \cdot g^t$.
\ENDFOR
\RETURN $G$.
\end{algorithmic}
\end{algorithm}

\section{Method}
\subsection{Overview}
Federated learning is highly susceptible to Byzantine attacks, particularly model poisoning attacks, where adversaries manipulate the updates of compromised clients to influence the aggregation process. These manipulations can alter both the direction and magnitude of model updates. To enhance robustness, it is essential to address both factors effectively. Our approach revolves around computing the cosine similarity between each local model update and the server model update. To filter out potentially malicious updates, we apply ReLU-clipped filtering, excluding clients whose cosine similarity falls below zero. To further mitigate the impact of data non-IID, we introduce a scoring mechanism based on the global model, ensuring a more balanced and fair weighting of client contributions during aggregation. After filtering, we normalize the magnitude of the remaining local model updates, reducing the influence of model poisoning attacks. Finally, we aggregate the filtered and normalized updates using a weighted averaging process to obtain the global model update.  The full workflow of our proposed FLTG framework is detailed in Algorithm~\ref{alg:fltg}, with the following Details section providing an in-depth explanation of its core techniques.

\begin{algorithm}[H]
\caption{FLTG}
\label{alg:fltg}
\begin{algorithmic}[1] 
\REQUIRE The server randomly selects $\eta$ clients $C_1, C_2, \ldots, C_\eta$ from the set $S = \{C_1, C_2, \ldots, C_n\}$, with their corresponding encrypted local model updates $g^t_{C_1}, g^t_{C_2}, \ldots, g^t_{C_\eta}$, server's model update $g^t_0$, the current round $t$, the global model update from the previous round $g^{t-1}$.
\ENSURE $Score_i, \bar{g}^t_i$.
\STATE $Score_i = Score'_i = Score''_i = 0$.
\STATE Initialize $S'$.
\FOR{$i = C_1, C_2, \ldots, C_\eta$}
    \STATE $Score'_i = \text{ReLU}\left(\frac{\langle {g}^t_i, g^t_0 \rangle}{\|{g}^t_i\| \cdot \|g^t_0\|}\right)$.
    \IF{$Score'_i > 0$}
        \STATE Add $i$ to the client subset $S'$.
    \ENDIF
\ENDFOR
\IF{$t == 1$}
    \FOR{$i \in S'$}
        \STATE $Score_i = Score'_i$.
    \ENDFOR
\ELSE
    \FOR{$i \in S'$}
        \STATE $Score''_i = \frac{\langle {g}^t_i, g^{t-1} \rangle}{\|{g}^t_i\| \cdot \|g^{t-1}\|}$.
    \ENDFOR
    \STATE $ref = \arg\min_{i \in S'} (Score''_i)$.
    \FOR{$i \in S'$}
        \STATE $Score_i = 1 - \frac{\langle {g}^t_i, g^t_{ref} \rangle}{\|{g}^t_i\| \cdot \|g^t_{ref}\|}$.
    \ENDFOR
\ENDIF
\FOR{$i \in S'$}
    \STATE $\bar{g}^t_i = \frac{\|g^t_0\|}{\|{g}^t_i\|} \cdot {g}^t_i$.
\ENDFOR
\RETURN $Score_i, \bar{g}^t_i$.
\end{algorithmic}
\end{algorithm}

\subsection{Details}
\begin{figure}[!ht]
    \centering
    \includegraphics[width=0.8\textwidth]{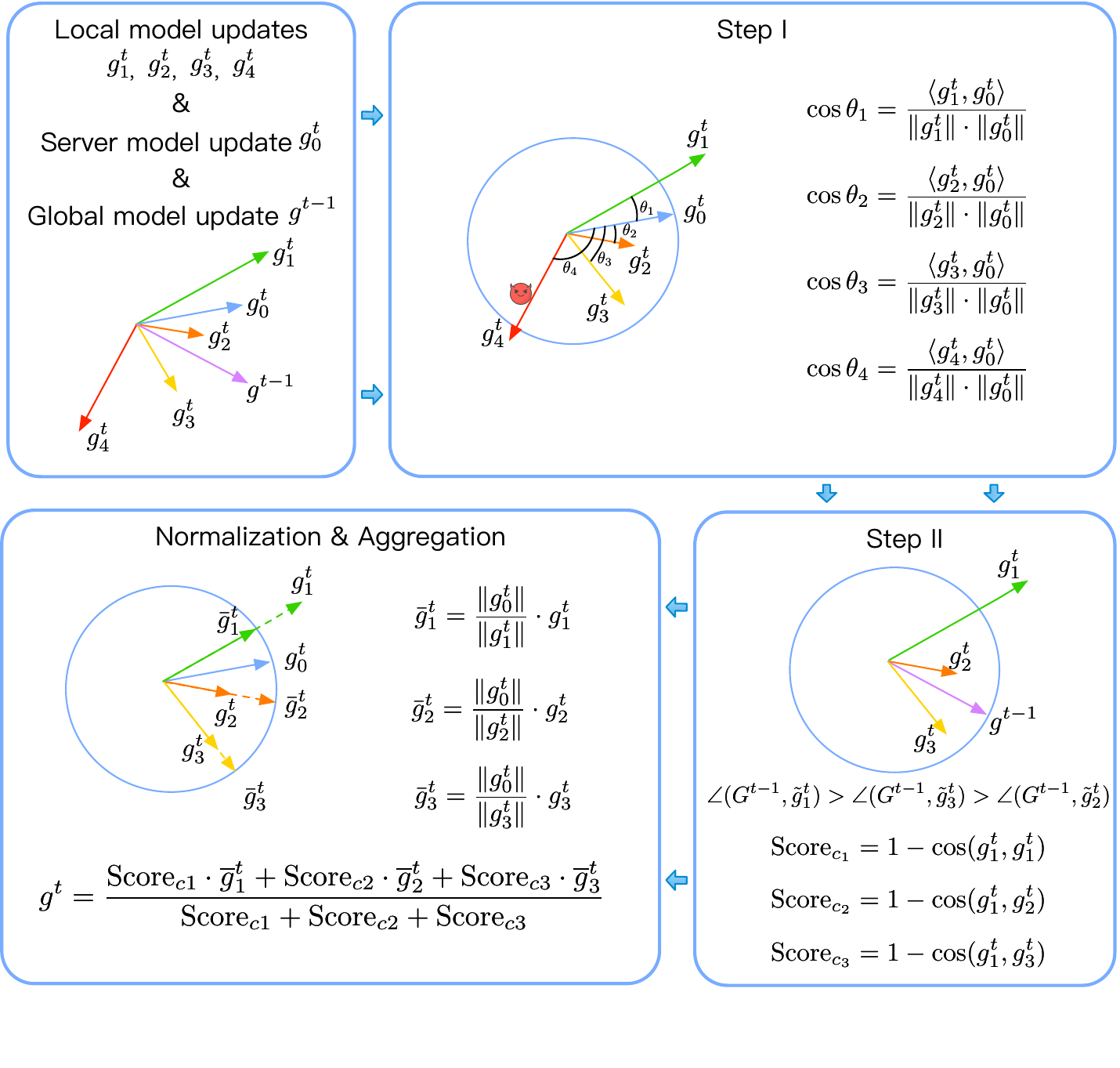}
    \caption{Illustration of FLTG.}
    \label{figures2}
\end{figure}

Our new aggregation rule implements a more reasonable scoring mechanism and stronger robustness.Fig.~\ref{figures2}. illustrates our aggregation rule.

\textbf{Cosine Similarity Filtering with ReLU Function:}  Attackers can manipulate the direction of malicious clients’ model updates to interfere with the global model update. To counter this, we treat the server model update as a trusted reference. Cosine similarity, a widely used metric for evaluating the angular relationship between two vectors, is employed to measure the directional alignment between each local model update and the server model update. If the cosine similarity is negative, the corresponding update could adversely affect the global aggregation. To prevent this, we apply the ReLU function to filter out such updates. The process is formally defined as follows:  

\begin{equation}  
\cos\theta_i = \frac{\langle \boldsymbol{g}_i^t, \boldsymbol{g}_0^t \rangle}{\|\boldsymbol{g}_i^t\| \cdot \|\boldsymbol{g}_0^t\|},  
\end{equation}  

where \( \boldsymbol{g}_0^t \) denotes the server model update at round \( t \), \( \| \cdot \| \) represents the \( \ell_2 \) norm of a vector, and \( \langle \boldsymbol{a}, \boldsymbol{b} \rangle \) is the inner product of vectors \( \boldsymbol{a} \) and \( \boldsymbol{b} \).

\noindent The ReLU function as folllows:
\begin{align}
\text{ReLU}(x) = 
\begin{cases} 
0 & \text{if } x < 0 \\
x & \text{if } x \geq 0 ,
\end{cases}
\end{align}

\textbf{Non-IID-Aware Weighting Mechanism:}
The non-IID nature of data distribution is a crucial factor in federated learning, as it can substantially affect model performance. To alleviate the impact of non-IID data and assign more appropriate aggregation weights to clients, we integrate the global model into the process, leveraging its comprehensive knowledge. Beyond handling non-IID challenges, we observed that simply filtering clients based on directional similarity does not entirely eliminate the risk of including malicious updates in the aggregation. To further mitigate their influence, we aim to assign lower aggregation weights to such clients. Our approach is structured as follows. All subsequent computations are performed on the subset \( S \) of clients that remain after applying the ReLU function. Within this subset, we first identify the client whose model update exhibits the lowest cosine similarity with the previous round’s global model. This client’s model update serves as a reference. We then compute the cosine similarity between this reference update and the updates from other clients in \( S \). The principle behind our weighting mechanism is that clients with greater similarity to the reference update should receive lower aggregation weights. The scoring function that determines these weights is defined as follows:  

\begin{equation}
\mathrm{Score}_{c_j} = 1 - \cos(\boldsymbol{g}_{\text{ref}}^t, \boldsymbol{g}_j^t),
\end{equation}  

where \( c_j \) represents a client in subset \( S \), \( \boldsymbol{g}_j^t \) denotes its corresponding encrypted model update at round \( t \), and \( \boldsymbol{g}_{\text{ref}}^t \) refers to the encrypted model update of the reference client in \( S \).

\textbf{Normalizing the magnitudes of local model updates:}
As mentioned earlier, model updates encompass both direction and magnitude. Attackers may also alter the magnitude of malicious clients’ updates to compromise the global model update. To counteract this, we apply normalization to each client’s model update, reducing the impact of such attacks. This normalization step also mitigates the influence of positive random scaling factors that clients may introduce to obscure their updates before submission while preserving the data’s utility.  

By leveraging the server model update as a reference, the normalization process projects all local updates onto the same hypersphere within the vector space. Specifically, each client’s update is scaled by its \( \ell_2 \) norm to form a unit vector. The normalization operation is defined as follows:  

\begin{equation}
\overline{\boldsymbol{g}}_j^t = \frac{\|\boldsymbol{g}_0^t\|}{\|\boldsymbol{g}_j^t\|} \cdot \boldsymbol{g}_j^t,
\end{equation}  

where \( \tilde{\boldsymbol{g}}_j^t \) represents the obfuscated model update of client \( j \) in subset \( S \) at round \( t \), \( \|\boldsymbol{g}_j^t\| \) is its corresponding \( \ell_2 \) norm, \( \|\boldsymbol{g}_0^t\| \) denotes the \( \ell_2 \) norm of the server model update, and \( \overline{\boldsymbol{g}}_j^t \) represents the normalized model update of client \( j \).

\textbf{Aggregating the local model updates:}  
In the final step, we calculate the weighted average of the normalized model updates from all clients to obtain the global model update. The aggregation formula is given by:  

\begin{equation}  
\boldsymbol{g} = \frac{1}{\sum_{j=1}^k Score_j} \sum_{j=1}^k Score_j \cdot \overline{\boldsymbol{g}}_j,  
\end{equation}  

where \( j \) denotes a client in the subset \( S \), consisting of \( k \) clients in total, \( Score_j \) represents the weight assigned to client \( j \) based on its contribution, and \( \overline{\boldsymbol{g}}_j \) is the normalized model update from client \( j \).

\section{Evaluation}
In this section, we present experimental results that evaluate the performance of our proposed method, FLTG, under existing poisoning attacks. We used the MNIST and CIFAR-10 datasets, with the number of clients set to 100 for both. The degree of non-IIDness in the data distribution among clients is represented by \( q \), where a larger value indicates a higher degree of non-IIDness. The size of the Root Dataset is 100, and its Bias Probability is \( p \), where a larger value indicates a greater degree of bias.

\begin{table*}[h]
    \centering
    \caption{Performance on MNIST under Standard Attacks.}
    \label{tab:shiyan1}
    \begin{subtable}{\textwidth}
        \centering
        \small
        \begin{adjustbox}{max width=\textwidth,scale=0.8}
        \begin{tabular}{l c c c c c c}
            \toprule
            & FedAvg & Krum & Trim-mean & Median & FLTrust & FLTG \\ 
            \midrule
            No attack & 0.9999 & 0.9374 & 0.9790 & 0.9790 & 1 & \textbf{1.0190} \\ 
            LF attack & 0.9790 & 0.9374 & 0.9895 & 0.9790 & 1 & \textbf{1.0169} \\ 
            Krum attack & 0.9374 & 0.0141 & 0.9686 & 0.9686 & 1 & \textbf{1.0115} \\ 
            Trim attack & 0.8749 & 0.9374 & 0.9061 & 0.9061 & 1 & \textbf{1.0080} \\ 
            Scaling attack & 1.0103/0.0000 & 0.9278/1.0000 & 0.9794/0.9900 & 0.9794/0.9900 & 1/1 & \textbf{1.0197}/\textbf{1.0012} \\ 
            Adaptive attack & 0.9583 & 0.9375 & 0.9271 & 0.9062 & 1 & \textbf{1.0184} \\ 
            \bottomrule
        \end{tabular}
        \end{adjustbox}
    \end{subtable}

    \small
    \begin{tabular}{p{0.95\textwidth}}  
        \textbf{Notes:} Scaling attack results are formatted as "Accuracy/Backdoor Defense Success Rate."
    \end{tabular}

\end{table*}

As shown in Table \ref{tab:shiyan1}, we use the same experimental settings in FLTrust\cite{2021FLTrust}: 5 types of standard attacks with a non-IID degree \(q=0.1\), a bias probability \(p=0.1\), and 20\% malicious clients. Note that we only reproduced FLTrust \cite{2021FLTrust} and use its results as the normalization bridge for performance comparison with other baseline methods. We calculate the ratio between the accuracy of other methods relative to FLTrust to ensure the fairness and consistency. We first calculated the ratio \( a \) as the accuracy of the locally reproduced FLTrust divided by the official FLTrust accuracy. Then, we multiplied the official performance values of each method by this ratio. Finally, we obtained the performance comparison of each method under the local hardware environment by dividing the result by the accuracy of the locally reproduced FLTrust. For instance, under scaling attacks, FLTrust achieved normalized accuracy and defense success rates of 1.0, while Median lagged by 2.06\% and 1\%, respectively. In contrast, FLTG outperformed FLTrust with 1.97\% higher accuracy and a 0.12\% improvement in defense rate. Notably, FLTG maintained near-optimal accuracy even in no-attack scenarios, demonstrating its compatibility with benign environments.

\begin{table*}[h]
    \centering
    \caption{Impact of Root Dataset Bias.}
    \label{tab:shiyan2}
    \begin{subtable}{\textwidth}
        \centering
        \small
        \begin{adjustbox}{max width=\textwidth}
        \begin{tabular}{l l c c c c c c}
            \toprule
            Attack & Method & 0.1 & 0.2 & 0.4 & 0.6 & 0.8 & 1.0 \\ 
            \midrule
            \multirow{2}{*}{No attack} & FLTrust & 0.9574  & 0.9531 & 0.9613 & 0.9552 & 0.9510 & 0.3546 \\  
            & FLTG & \textbf{0.9681}  & \textbf{0.9592} & \textbf{0.9640} & \textbf{0.9652} & \textbf{0.9620} & \textbf{0.9196} \\ 
            \midrule
            \multirow{2}{*}{LF attack} & FLTrust & 0.9564 & 0.9587 & 0.9595 & 0.9526 & 0.9562 & 0.0980 \\  
            & FLTG & \textbf{0.9661} & \textbf{0.9671} & \textbf{0.9675} & \textbf{0.9692} & \textbf{0.9640} & \textbf{0.1261} \\ 
            \midrule
            \multirow{2}{*}{Krum attack} & FLTrust & 0.9457 & 0.9445 & 0.9421 & 0.9101 & 0.9384 & \textbf{0.1330} \\  
            & FLTG & \textbf{0.9509}  & \textbf{0.9532} & \textbf{0.9504} & \textbf{0.9107} & \textbf{0.9429} & 0.0980 \\ 
            \midrule
            \multirow{2}{*}{Trim attack} & FLTrust & 0.9541 & 0.9404 & 0.9452 & 0.9343 & 0.9316 & 0.0980 \\  
            & FLTG & \textbf{0.9576}  & \textbf{0.9550} & \textbf{0.9502} & \textbf{0.9524} & \textbf{0.9515} & 0.0980 \\ 
            \midrule
            \multirow{2}{*}{Scaling attack} & FLTrust & 0.9558 / 0.0072 & 0.9494 / 0.0093 & 0.9577 / 0.0194 & 0.9501 / 0.9020 & 0.9535 / 0.9020 & 0.0980 / 0.9020 \\  
            & FLTG & \textbf{0.9688} / \textbf{0.0060} & \textbf{0.9565} / \textbf{0.0077} & \textbf{0.9651} / \textbf{0.0139} & \textbf{0.9612} / \textbf{0.0285} & \textbf{0.9683} / \textbf{0.0465} & 0.0980 / 0.9020 \\ 
            \midrule
            \multirow{2}{*}{Adaptive attack} & FLTrust & 0.9452  & 0.9275 & 0.9366 & 0.0980 & 0.0980 & 0.0980 \\  
            & FLTG & \textbf{0.9573}  & \textbf{0.9500} & \textbf{0.9536} & \textbf{0.9434} & \textbf{0.9369} & 0.0980 \\ 
            \bottomrule
        \end{tabular}
        \end{adjustbox}
    \end{subtable}

\end{table*}

In Table \ref{tab:shiyan2}, to investigate FLTG's tolerance to root dataset bias, we set \( q = 0.1 \) and varied the root dataset bias probability \( p \) from 0.1 to 1.0 FLTG consistently outperformed FLTrust across all bias levels. In the no-attack scenario with extreme bias (\( p = 1.0 \)), FLTrust's accuracy dropped significantly to 0.3546, while FLTG maintained an accuracy of 0.9196. This demonstrates FLTG's ability to effectively mitigate the impact of data heterogeneity, even when the root dataset is highly skewed.

\begin{figure}[htbp]
    \centering
    \begin{subfigure}{0.33\textwidth} 
        \includegraphics[width=\linewidth]{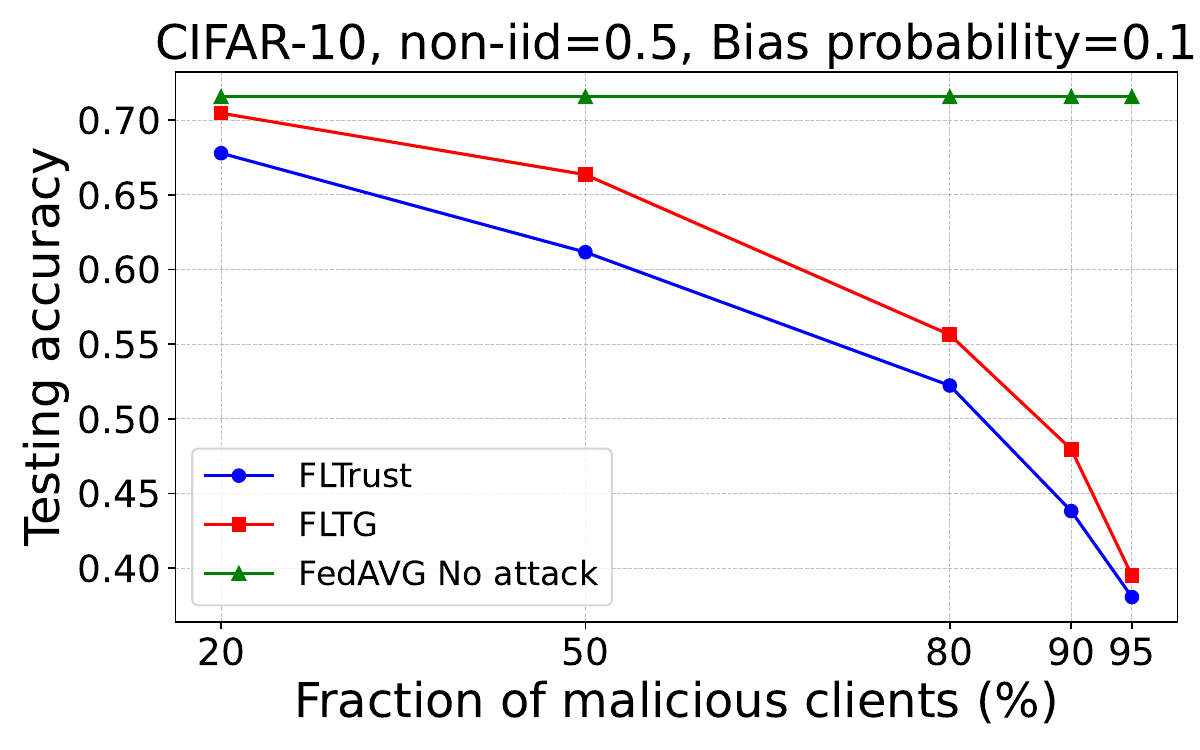}
        \caption{Bias probability=0.1}
        \label{fig:chart1}
    \end{subfigure}%
    \hfill
    \begin{subfigure}{0.33\textwidth} 
        \includegraphics[width=\linewidth]{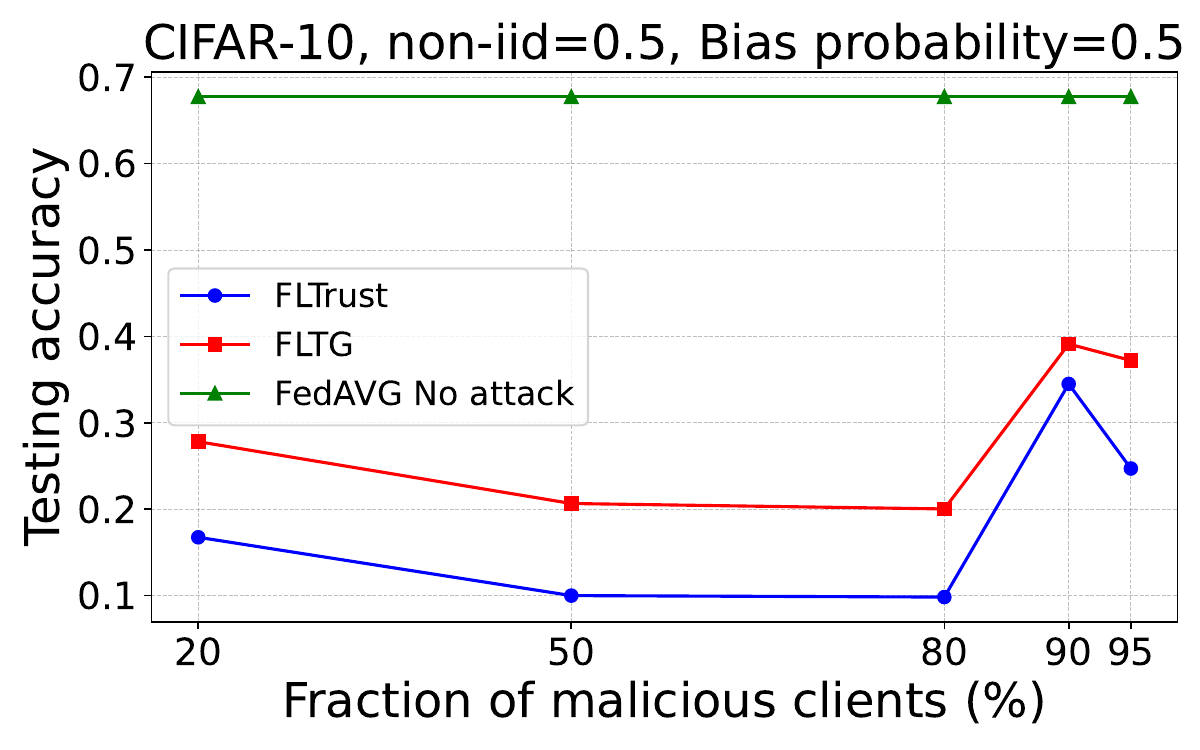}
        \caption{Bias probability=0.5}
        \label{fig:chart2}
    \end{subfigure}%
    \hfill
    \begin{subfigure}{0.33\textwidth} 
        \includegraphics[width=\linewidth]{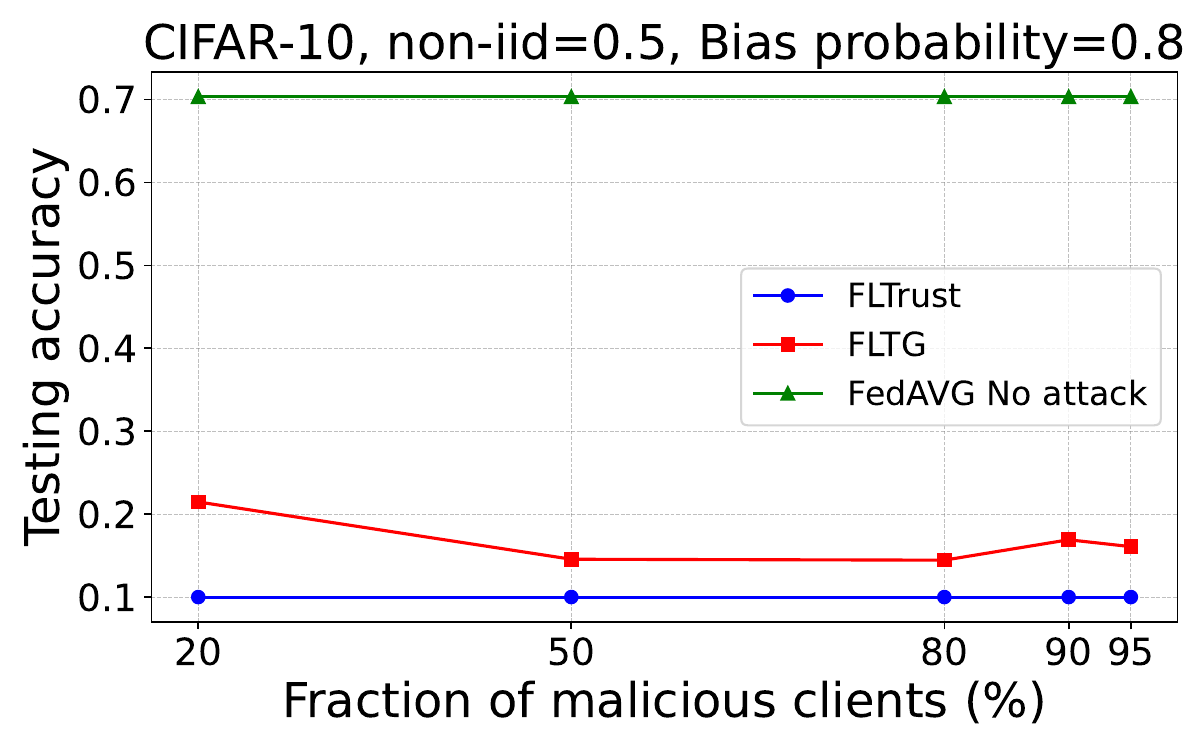}
        \caption{Bias probability=0.8}
        \label{fig:chart3}
    \end{subfigure}
    \caption{Impact of Root Dataset Bias Probability and Malicious Client Ratio on Model Robustness under Adaptive Poisoning Attack.}
    \label{fig:charts1}
\end{figure}

In Figure \ref{fig:charts1}, we have replaced the dataset with the more complex CIFAR-10 dataset under higher non-IID (q=0.5), and employed the stronger model poisoning attack, Min-max\cite{shejwalkar2021manipulating}. The figure shows a comparison of FLTG's performance against FLTrust under varying bias probabilities and fractions of malicious clients, along with FedAvg under no-attack conditions. The experiments demonstrate that FLTG outperforms FLTrust in terms of both bias probability tolerance and resilience to varying proportions of malicious clients. Specifically, at a bias probability of 0.8, where conditions are most challenging, FLTG still manages to maintain a relatively higher testing accuracy than FLTrust, indicating its robustness against high levels of dataset bias and malicious client influence. These results highlight the effectiveness of FLTG in maintaining performance under more complex and adversarial conditions.

\begin{figure}[htbp]
    \centering
    \begin{subfigure}{0.33\textwidth}
        \includegraphics[width=\linewidth]{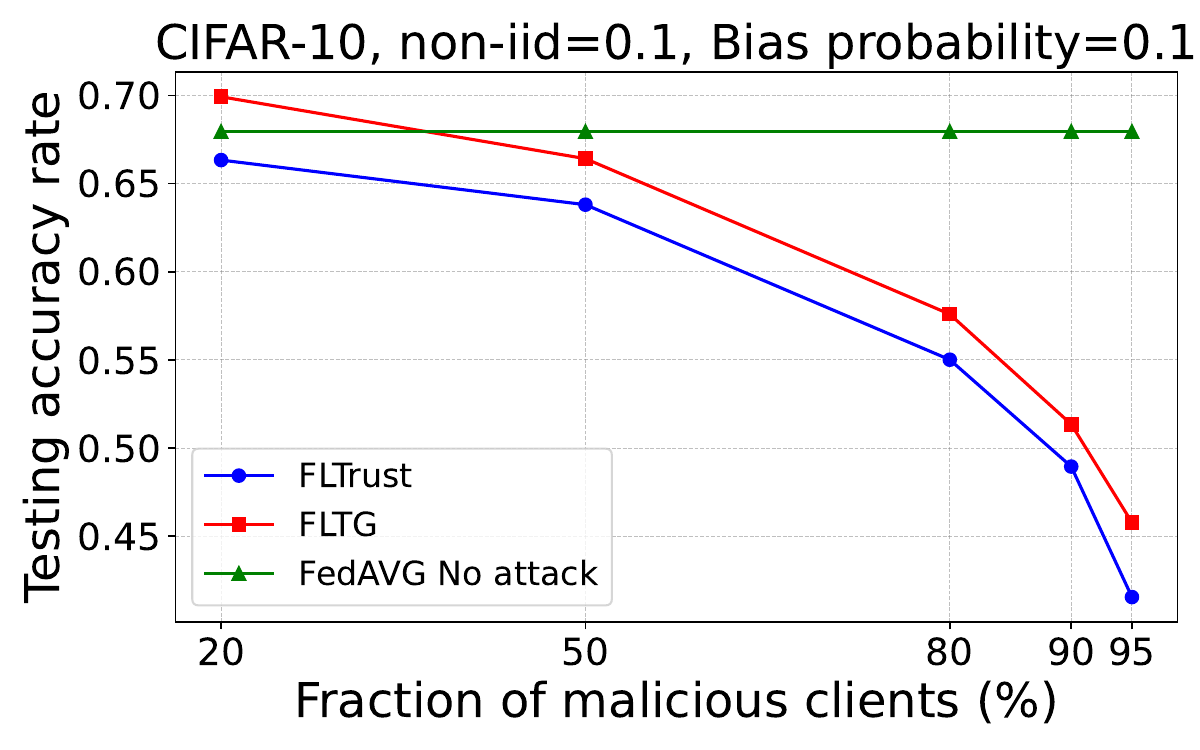}
        \caption{Non-iid=0.1}
        \label{fig:chart4}
    \end{subfigure}%
    \hfill
    \begin{subfigure}{0.33\textwidth}
        \includegraphics[width=\linewidth]{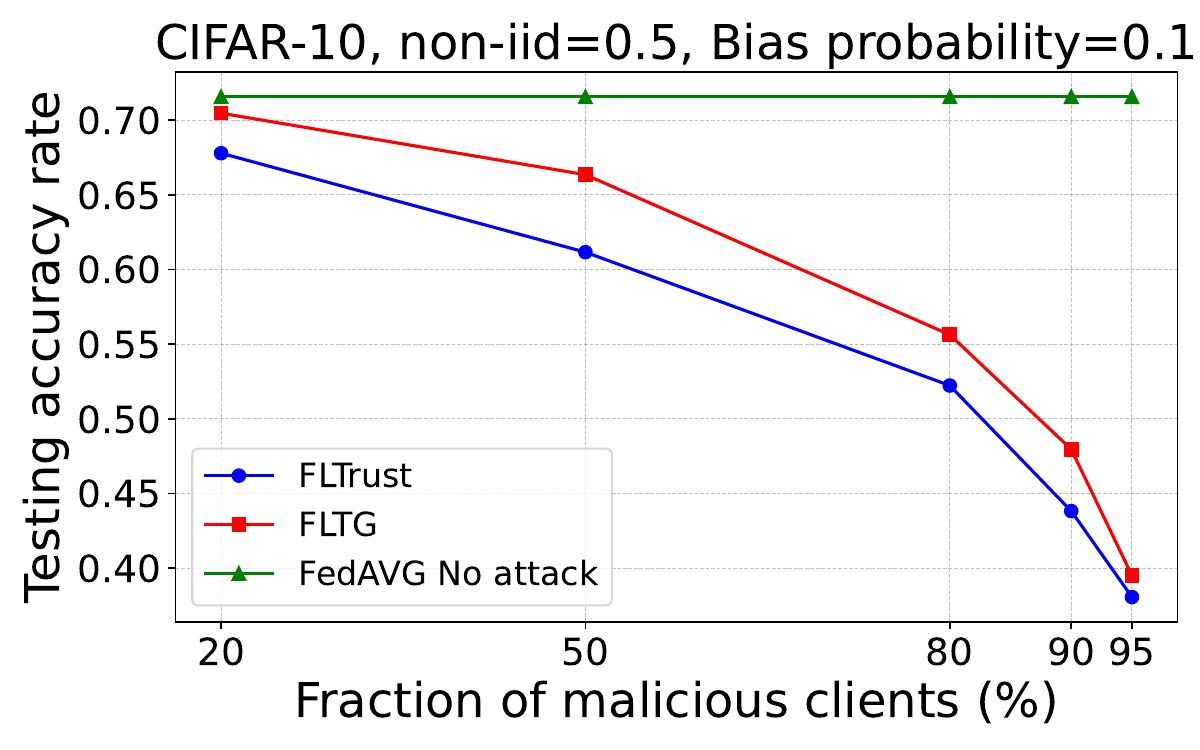}
        \caption{Non-iid=0.5}
        \label{fig:chart5}
    \end{subfigure}%
    \hfill
    \begin{subfigure}{0.33\textwidth}
        \includegraphics[width=\linewidth]{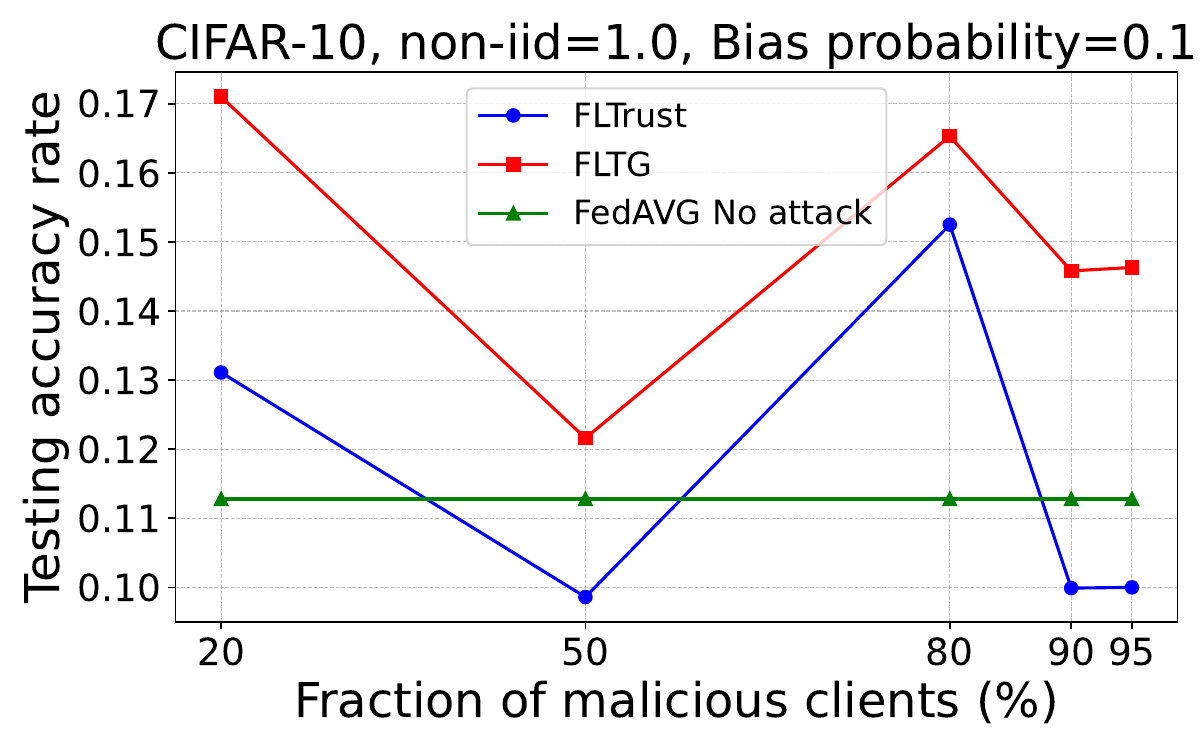}
        \caption{Non-iid=1.0}
        \label{fig:chart6}
    \end{subfigure}
    \caption{Impact of Client Non-IID Degree and Malicious Client Ratio on Model Robustness under Adaptive Poisoning Attack.}
    \label{fig:charts2}
\end{figure}

Finally, in Figure \ref{fig:charts2}, we also use the CIFAR-10 dataset. We fix p=0.1 and use the Min-max attack to study the performance of FLTG compared to FLTrust under varying levels of non-IID and fractions of malicious clients, as well as the performance of FedAvg under no-attack conditions. Specifically, in subplot (a) of Figure \ref{fig:charts2}, where the non-IID level is set to 0.1, FLTG not only surpasses FLTrust but also achieves a higher testing accuracy rate than FedAvg in no-attack scenarios, even at a low fraction of malicious clients, such as 20\%. This indicates that even at a relatively low level of non-IID distribution and under attack by a relatively small proportion of malicious clients, FLTG is still able to effectively maintain high performance.

\section{Conclusion and Future Work}
This paper introduces FLTG, a new aggregation algorithm for defending against Byzantine attacks in federated learning. Using cosine similarity screening and model update normalization, FLTG enhances model convergence and resilience to poisoning attacks. Future work will focus on improving privacy and efficiency in federated learning, aiming for a balance between privacy, accuracy, and computational cost. We will also conduct scalability evaluations in large-scale federated learning environments and investigate how performance changes as the number of clients increases, in order to broaden its applicability and enhance the practical relevance of the paper.

\subsubsection{Acknowledgments.}This research was supported in part by the National Social Science Foundation Youth Program (Grant No. 24CGL104), Hunan Provincial Natural Science Foundation of China (No. 2023JJ40237), and the Youth Program of Hunan Provincial Department of Education (Grant No. 23B0598, No. 22B0648).

\bibliographystyle{spmpsci_unsrt}
\bibliography{reference}

\end{document}